\documentclass[pra,aps,showpacs,superscriptaddress, reprint]{revtex4-1}
\usepackage{amsmath}

\usepackage{graphicx}
\graphicspath{{figures/}}
\usepackage{epsfig}
\usepackage{float}
\usepackage{tikz}
\usetikzlibrary{calc, intersections, decorations.markings}

\def\ket#1{|\,#1 \,\rangle}
\def\bra#1{\langle \, #1 \,|}

\def\vop#1{\mathbf{#1}}

\def\kHz{\,\mathrm{kHz}}

\def\eref#1{Eq.~(\ref{#1})}
\def\fref#1{Fig.~\ref{#1}}

\begin{document}
\title{Dicke model simulation via cavity-assisted Raman transitions}
\author{Zhang Zhiqiang}
\email{e0000155@u.nus.edu}
\affiliation{Centre for Quantum Technologies, 3 Science Drive 2, 117543 Singapore}
\author{Chern Hui Lee}
\affiliation{Centre for Quantum Technologies, 3 Science Drive 2, 117543 Singapore}
\author{Ravi Kumar}
\affiliation{Centre for Quantum Technologies, 3 Science Drive 2, 117543 Singapore}
\author{K. J. Arnold}
\affiliation{Centre for Quantum Technologies, 3 Science Drive 2, 117543 Singapore}
\author{Stuart J. Masson}
\affiliation{Dodd-Walls Centre for Photonics and Quantum Technologies, Department of Physics, University of Auckland, Private Bag 92019, Auckland, New Zealand}
\author{A. L. Grimsmo}
\affiliation{ARC Centre of Excellence for Engineered Quantum Systems, School of Physics, The University of Sydney, Sydney, NSW 2006, Australia}
\author{A. S. Parkins}
\affiliation{Dodd-Walls Centre for Photonics and Quantum Technologies, Department of Physics, University of Auckland, Private Bag 92019, Auckland, New Zealand}
\author{M. D. Barrett}
\affiliation{Centre for Quantum Technologies, 3 Science Drive 2, 117543 Singapore}
\affiliation{ Department of Physics,National University of Singapore,  2 Science Drive 3, 117551 Singapore}
\begin{abstract}
The Dicke model is of fundamental importance in quantum mechanics for understanding the collective behaviour of atoms coupled to a single electromagnetic mode.
In this paper, we demonstrate a Dicke-model simulation using cavity-assisted Raman transitions in a configuration using counter-propagating laser beams. The observations indicate that motional effects should be included to fully account for the results and these results are contrasted with the experiments using single-beam and co-propagating configurations. A theoretical description is given that accounts for the beam geometries used in the experiments and indicates the potential role of motional effects.  In particular a model is given that highlights the influence of Doppler broadening on the observed thresholds.
\end{abstract}
\maketitle

\section{Introduction}

 An ensemble of $N$ two-level atoms coupled to a single mode of an electromagnetic field is an important problem in quantum electrodynamics (QED) and continues to be an active avenue of research, not only in cavity QED \cite{Baumann2010, bastidas2012nonequilibrium, Zhiqiang2016, kirton2017suppressing} but also equivalent realizations in circuit QED with artificial atoms \cite{niemczyk2010circuit, viehmann2011superradiant, zhang2014quantum, zhu2016quantum}.  A simple model of this interaction is provided by the Dicke Hamiltonian \cite{dicke1954coherence,dimer2007proposed} given by
\begin{equation}
H=\omega a^\dagger a+\omega_0 J_\mathrm{z}+\frac{\lambda}{\sqrt{N}}(a^\dagger+a)(J_+ + J_-)
\end{equation}
where $\omega_0$ represents the frequency splitting of the two levels, $\omega$ is the frequency of the electromagnetic field, and $\lambda$ is the interaction strength.   The operator $a$ is the usual annihilation operator for the field, and the operators $J_\mathrm{\pm}$ and $J_\mathrm{z}$ are the collective atomic operators which satisfy the usual angular momentum commutation relations $[J_+, J_-]=2J_\mathrm{z}$ and $[J_\mathrm{z}, J_\mathrm{\pm}]=\pm J_\mathrm{\pm}$.  

Experimental realizations of this system have utilized cavity-assisted Raman transitions between momentum states of a Bose$-$Einstein condensate (BEC) \cite{Klinder2015, Baumann2010} or internal states of thermal atoms \cite{dimer2007proposed,baden2014realization, *baden2017erratum,Zhiqiang2016}.
The implementation given in \cite{baden2014realization} was based on a scheme derived from the proposal in \cite{dimer2007proposed}.  Since that work, we have made a number of improvements to the experiment, including the use of a field programmable gate array (FPGA) to provide real-time control of the atom number based on non-destructive measurements of the cavity dispersion, and more independent control of both the coupling beams.  These improvements allow cleaner data collection that is less dependent on post-selection.  It has also enabled us to explore the imbalanced driving case using an implementation based on a collection of spin-1 atoms \cite{Zhiqiang2016}.  

In this work we give a detailed account of a Dicke-model simulation using cavity-assisted Raman transitions along with a discussion of relevant theory, which follows the original proposal in \cite{dimer2007proposed} but allows for more general beam geometries as used in the experiments. The beam geometry has a significant influence on the potential effects of motion and the validity of a Dicke model interpretation. Counter-  and co-propagating beam geometries are demonstrated experimentally and contrasted with a single beam configuration. Observed thresholds agree with theoretical predictions that includes the influence Doppler broadening.  This work provides useful information for understanding interaction of light and an atomic ensemble coupled to a single electromagnetic mode and the calculations and experiment results will provide a helpful reference for future experimental work.

\section{Experimental Set Up}
\begin{figure}
 \includegraphics[width=0.4\textwidth]{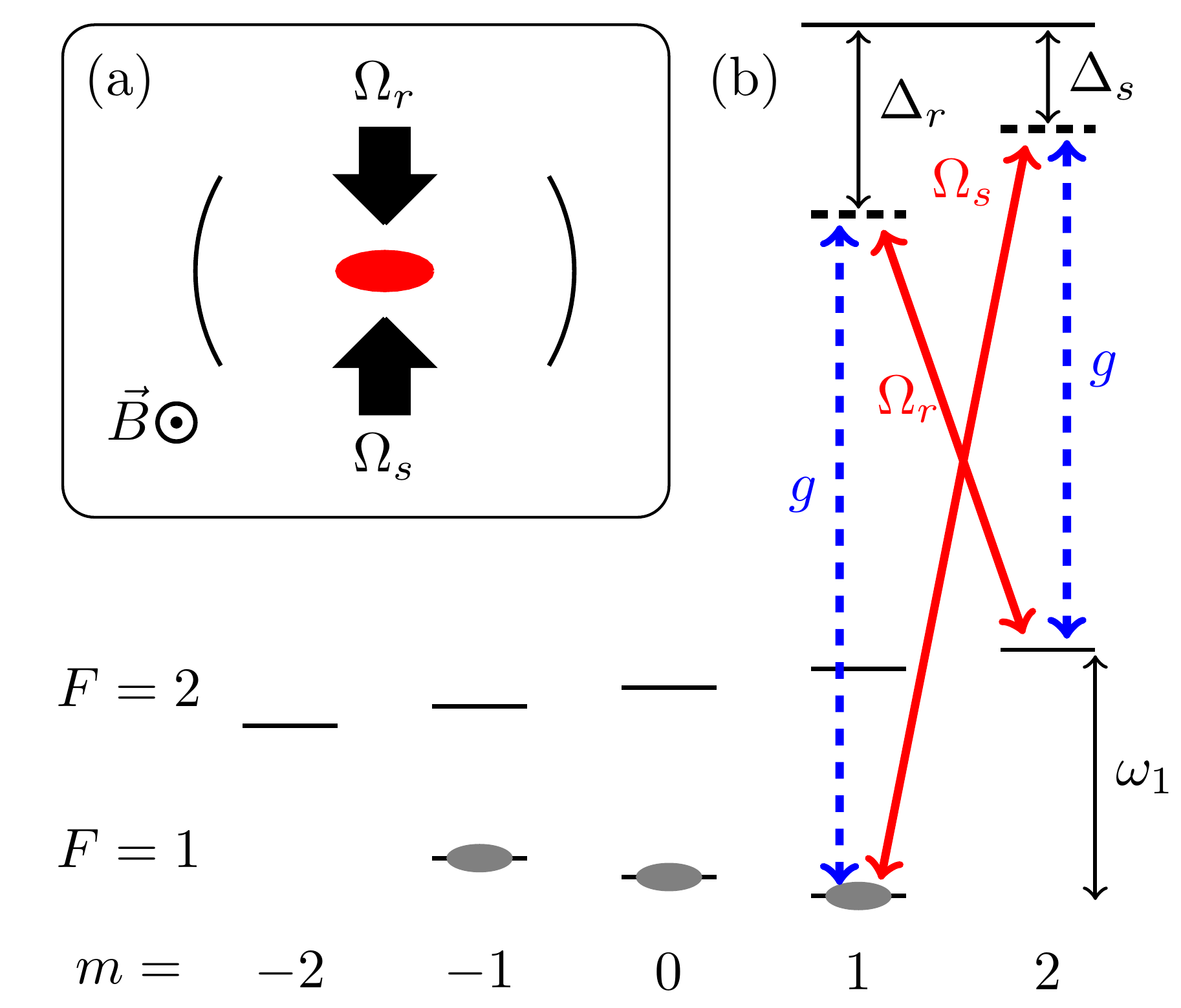}
 \caption{(a) Schematic representation of the counter-propagating probe beam geometry, and (b) atomic level structure for Dicke model implementation. Red lines are classical beam couplings and dashed blues lines are cavity mode couplings. The frequency $\omega_1$ is the separation of the two states relevant to the Dicke model inclusive of the ground state hyper-fine splitting and Zeeman shifts. Note that as defined in the text, the detunings $\Delta_\mathrm{r}$ and $\Delta_\mathrm{s}$ would have negative values as drawn in the figure. }
  \label{fig:transitiondiag}
\end{figure}

The experimental setup is similar to the one described elsewhere \cite{arnold2012self, Zhiqiang2016}.  An ensemble of $^{87}$Rb atoms are trapped within an optical cavity using an intra-cavity $1560\,\mathrm{nm}$ optical lattice. The cavity is detuned from the $^2\mathrm{S}_{1/2}$ to $^2\mathrm{P}_{3/2}$ transition by $\Delta=-2\pi\times 127\,\mathrm{GHz}$. Relative to this transition, the cavity parameters are $(g, \kappa, \gamma_\mathrm{a}) = 2\pi \times(1.1,0.1, 3)$\,MHz, where  $g$ is the single atom-cavity coupling constant for the $\ket{F=2,m_\mathrm{F} = 2}$ to  $\ket{F' =3, m_\mathrm{F'} = 3}$ cycling transition; $\kappa$ and $\gamma_\mathrm{a}$ are half-width-half-maximum linewidth of the cavity and the atomic dipole decay rate, respectively.  The wavelength of the 1560\,nm optical lattice is exactly twice the cavity resonance near the $^2\mathrm{S}_{1/2}$ to $^2\mathrm{P}_{3/2}$ transition, thus the atoms are trapped very close to alternate antinodes of the cavity field, maximizing the atom-cavity coupling.  The atoms are driven transverse to the cavity by two laser beams which are approximately equal in power.  As with the cavity coupling, the coupling strengths $\Omega_\mathrm{s}$ and $\Omega_\mathrm{r}$ are for the $\ket{F=2,m_\mathrm{F} = 2}$ to  $\ket{F' =3, m_\mathrm{F'} = 3}$ cycling transition. Each beam drives a cavity-assisted Raman transition between the two hyperfine ground states as depicted in \fref{fig:transitiondiag}. The beams are linearly polarized with a polarization orthogonal to a magnetic field that defines the quantization axis.  The cavity output is detected using either a single-photon counting module (SPCM) for maximum sensitivity or optical hetrodyne detection for larger dynamic range.
\section{Theory}
\label{Sect:theory}
Derivation of the Hamiltonian follows the treatment given in \cite{dimer2007proposed} while retaining the phases of the driving fields.  After adiabatic elimination of the excited states, the Hamiltonian for the experimental setup shown in \fref{fig:transitiondiag} is given by
\begin{equation}
 H = \omega \; a^\dagger a + \omega_0 \; J_\mathrm{z} + \frac{\delta}{N} a^\dagger a \; J_\mathrm{z} + H_\mathrm{R}
\end{equation}
where 
\begin{multline}
H_\mathrm{R} = \frac{\sqrt{3}}{12\Delta_\mathrm{r}}\sum_{j=1}^{N} \left( g(\vop{r}_j) \; \Omega_\mathrm{r} \; e^{-i (\vop{k}_r \cdot \vop{r}_j+\theta_\mathrm{r})} a J_{+,j}+ \text{h.c.}\right) \\
+ \frac{\sqrt{3}}{12\Delta_\mathrm{s}}\sum_{j=1}^{N} \left( g(\vop{r}_j) \; \Omega_\mathrm{s} \; e^{i (\vop{k}_\mathrm{s} \cdot \vop{r}_j+\theta_\mathrm{s})} a^\dagger J_{+,j} + \text{h.c.}\right),
\end{multline}
\begin{multline}
  \omega_0 = \omega_1-\frac{1}{2}\left(\omega_\mathrm{s}-\omega_\mathrm{r}\right)\\
  +\frac{1}{6} \left\{\left( \frac{\Omega_\mathrm{r}^2}{\Delta_\mathrm{r}}  - \frac{\Omega_\mathrm{r}^2}{\Delta_\mathrm{r} - \omega_1}\right)-\left( \frac{\Omega_\mathrm{s}^2}{\Delta_\mathrm{s}}-\frac{\Omega_\mathrm{s}^2}{\Delta_\mathrm{s}+\omega_1}\right) \right\},
\end{multline}
\begin{equation}
  \omega =\omega_c-\frac{1}{2}\left(\omega_\mathrm{r}+\omega_\mathrm{s}\right)+\frac{N}{3} \left(\frac{\langle g^2\rangle}{\Delta_\mathrm{s}} + \frac{\langle g^2\rangle}{\Delta_\mathrm{r}}\right)
\end{equation}
and
\begin{equation}
\delta = \frac{2 N}{3} \left(\frac{\langle g^2\rangle}{\Delta_\mathrm{s}} - \frac{\langle g^2\rangle}{\Delta_\mathrm{r}}\right).
\end{equation}
In these expressions $\langle \cdot \rangle$ is an average over the spatial distribution of atoms and $\omega_1$ is the Zeeman shifted hyperfine splitting between the states of interest.  With  the definitions
\begin{eqnarray}
\ket{0}&\!\equiv \ket{F=1,m_\mathrm{F}=1}\\
\ket{1}&\,\equiv \ket{F=2,m_\mathrm{F}=2}, 
\end{eqnarray}
the single particle angular momentum operators $J_{\pm,j}$ and $J_{z,j}$ are defined by
\begin{equation}
J_{+,j}=\ket{1_j}\bra{0_j},\quad J_{-,j}=\ket{0_j}\bra{1_j},
\end{equation}
and
\begin{equation}
J_{z,j}=\frac{1}{2}\left(\ket{1_j}\bra{1_j}-\ket{0_j}\bra{0_j}\right),
\end{equation}
with collective counterparts
\begin{equation}
J_{\pm}=\sum_j J_{\pm,j},\quad J_\mathrm{z}=\sum_j J_{z,j}.
\end{equation}
For a suitable choice of $\theta_\mathrm{r,s}$, $\Omega_\mathrm{r,s}$ can be assumed real. However, the dependence on $\theta_\mathrm{r}$ and $\theta_\mathrm{s}$ can be removed using the unitary transform
\begin{equation}
U_\theta=\exp\left(i \frac{\theta_\mathrm{r}-\theta_\mathrm{s}}{2} J_\mathrm{z}\right)\exp\left(-i \frac{\theta_\mathrm{r}+\theta_\mathrm{s}}{2} a^\dagger a\right),
\end{equation}
assuming these do not have significant temporal variation over the timescale of an experiment. 

As the atoms are confined at alternate antinodes of the cavity mode, $g(\vop{r}_j)$ is approximately constant and we approximate it by its thermally averaged value, $\langle g\rangle$.  For counter-propagating beams, $\vop{k}_\mathrm{r}\approx-\vop{k}_\mathrm{s}\equiv\vop{k}$ and the phase terms $e^{\pm i \vop{k}\cdot\vop{r}_j}$ can be removed from $H_\mathrm{R}$ using the unitary transformation
\begin{equation}
U_\mathbf{k}=\prod_j\exp\left[i (\vop{k}\cdot\vop{r}_j) J_{z,j}\right].
\label{motion}
\end{equation}
This then leads to the Hamiltonian
\begin{multline}
 H = \omega a^\dagger a + \omega_0 J_\mathrm{z} + \frac{\delta}{N} a^\dagger a J_\mathrm{z} \\
 + \frac{\lambda_\mathrm{r}}{\sqrt{N}}(a J_++a^\dagger J_-)
 +\frac{\lambda_\mathrm{s}}{\sqrt{N}}(a J_-+a^\dagger J_+)+H_1
 \label{Eq:Hamiltonian}
\end{multline}
where 
\begin{equation}
\lambda_\mathrm{r}=\frac{\sqrt{3N}}{12}\frac{\Omega_\mathrm{r}\langle g\rangle}{\Delta_\mathrm{r}},\quad \lambda_\mathrm{s}=\frac{\sqrt{3N}}{12}\frac{\Omega_\mathrm{s} \langle g\rangle}{\Delta_\mathrm{s}}.
\end{equation}
In carrying out this last unitary transformation, we have treated the positions as fixed numbers which effectively ignores motional effects.  Including the motion adds a term
\begin{equation}
 H_1=\omega_\mathrm{T}\sum_j  b_j^\dagger b_j+\sum_j (\mathbf{k}\cdot\mathbf{v}_j) J_{z,j},
 \label{eq:Doppler}
\end{equation}
to the Hamiltonian where $\omega_\mathrm{T}$ denotes the trap frequency associated with the harmonic confinement along the propagation direction of the lasers, and $b_j$ the associated ladder operator for the $j^\mathrm{th}$ atom.  The last term in the expression for $H_1$ arises from the transformation given in Eq.~\ref{motion} and it reflects the sensitivity of $\omega_0$ to Doppler shifts in the counter-propagating configuration.  Interpretation as a simple, idealized Dicke model thus implicitly neglects motional effects and requires the classical beams to be counter-propagating.

The parameters $\omega$ and $\omega_0$ are specified in terms of the mean and difference of the laser frequencies, which can be conveniently referenced, respectively, to the cavity resonance and the hyperfine splitting between the two states of interest.  They can equally be expressed in terms of the detuning of each beam from the Raman resonance with the cavity.  For this purpose, we define
\begin{equation}
\omega_\mathrm{d}= \frac{N}{3} \left(\frac{\langle g^2\rangle}{\Delta_\mathrm{s}} + \frac{\langle g^2\rangle}{\Delta_\mathrm{r}}\right)
\end{equation}
and
\begin{multline}
\Delta\omega_\mathrm{ss}= \frac{1}{6} \Bigg\{\left( \frac{\Omega_\mathrm{r}^2}{\Delta_\mathrm{r}}  - \frac{\Omega_\mathrm{r}^2}{\Delta_\mathrm{r} - \omega_1}\right)\\-\left( \frac{\Omega_\mathrm{s}^2}{\Delta_\mathrm{s}}-\frac{\Omega_\mathrm{s}^2}{\Delta_\mathrm{s}+\omega_1}\right)\Bigg\}
\end{multline}
and we may write
\begin{subequations}
\begin{eqnarray}
\delta_\mathrm{cr} &\equiv&\omega_0-\omega=\omega_\mathrm{r}-\omega_c+\omega_1-\omega_\mathrm{d}+\Delta\omega_\mathrm{ss}\\
\delta_\mathrm{cs} &\equiv&-(\omega_0+\omega)=\omega_\mathrm{s}-\omega_c-\omega_1-\omega_\mathrm{d}-\Delta\omega_\mathrm{ss}
\end{eqnarray}
\end{subequations}
In the limit that $\delta=0$, these are simply the detunings of each beam from Raman resonance with the cavity, properly accounting for cavity dispersion and AC stark shifts.   For $\delta\neq 0$, the detunings from Raman resonance become a dynamical quantity as the population moves from one state to the other.   

Since $|\Delta_\mathrm{r,s}|\gg\omega_1$, it is convenient to write $\Delta_\mathrm{r}=\Delta-\omega_1/2$ and $\Delta_\mathrm{s}=\Delta+\omega_1/2$.  Then, taking $\Omega_\mathrm{r}\approx\Omega_\mathrm{s}=\Omega$, we have 
\begin{equation}
\omega_\mathrm{d}=\frac{2}{3}\frac{N \langle g^2\rangle}{\Delta}, \quad \delta=-\omega_\mathrm{d} \frac{\omega_1}{\Delta},\quad \Delta\omega_\mathrm{ss}=-\frac{1}{3}\frac{\Omega^2}{\Delta^2}\omega_1
\end{equation}
correct to first order in $\omega_1/\Delta$.  In all cases, $N$ represents the number of atoms within the two-level system of interest.  Atoms outside this subspace simply add an additional dispersive shift in the definition of $\omega$.

\begin{figure*}[ht]
 \includegraphics{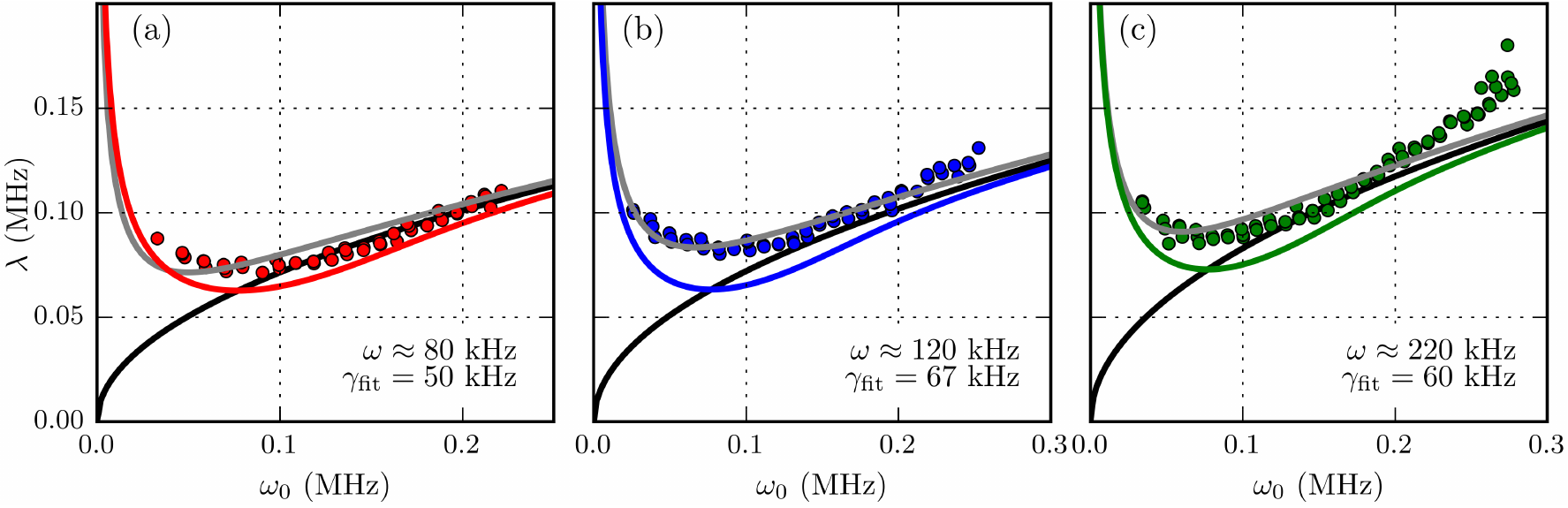}
 \caption{Measured critical coupling as a function of $\omega_0$  for selected fixed values of $\omega$ (a-c). The solid black lines are an \emph{an initio} calculation of the threshold from \eref{Eq:thresholdDimer}. Gray curves are fit to the experimental data using \eref{Eq:threshold} with $\langle \sigma_\mathrm{z} \rangle = -0.5$ and $\gamma$ as a free parameter. For  experimental data the trap depth is fixed at $219\,\mathrm{\mu K}$. Considering the equivalent rms Doppler broadening of $\gamma_\mathrm{d}=2\pi \times 59\,\mathrm{kHz}$ colored solid curves are obtained using \eref{Eq:threshold_inhomogeneous}. }
\label{threshold1}
\end{figure*}
\begin{figure*}
 \includegraphics{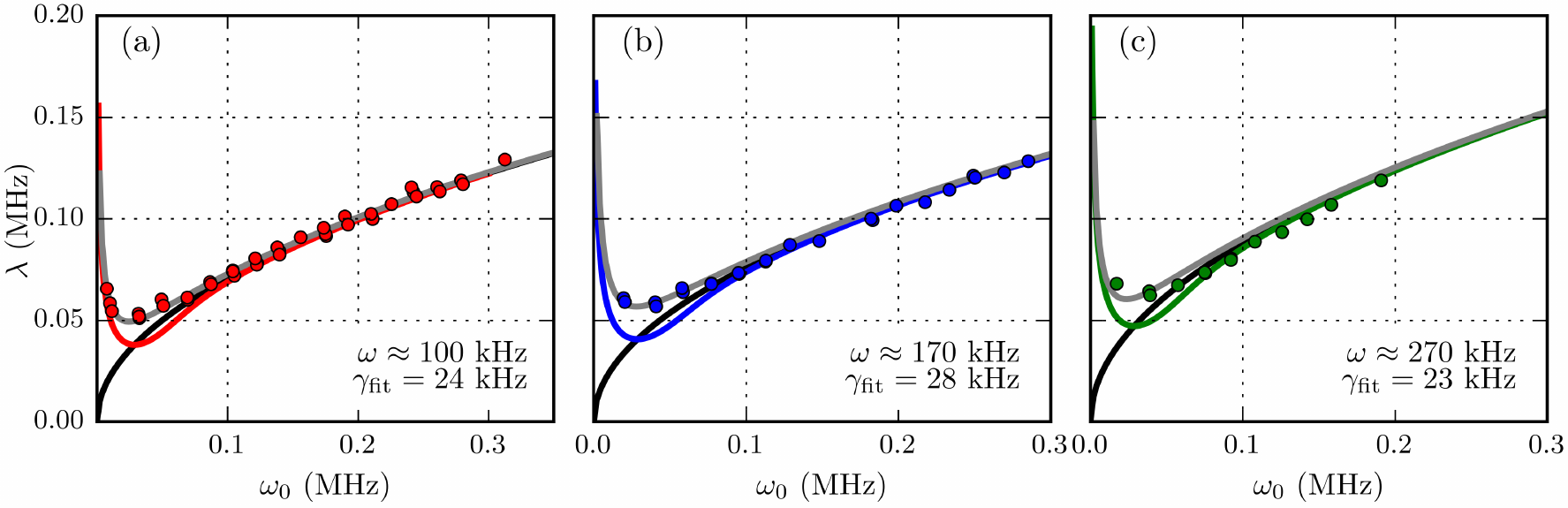}
\caption{Same as \fref{threshold1} except with trap depth fixed at $31\,\mathrm{\mu K}$, with the equivalent rms Doppler broadening of $2\pi \times  22\,\mathrm{kHz}$}.
\label{threshold2}
\end{figure*}

\section{Experimental Implementation}
An experiment starts by preparing a set number of atoms in the $F=1$ ground-state manifold with a well-defined temperature which is achieved as follows.  First, a magneto-optical trap (MOT) is formed 15\,mm above the cavity.  The atoms are then pumped into the $F=1$ hyperfine manifold and transferred to a single-beam 1064\,nm dipole trap that is overlapped with the MOT.  Typically, around $5\times10^6$ atoms are loaded into the dipole trap.  Using a motorized translation stage, the beam is then moved down 15\,mm over one second to bring the atoms into the cavity. The power of the 1064\,nm beam is then adiabatically lowered in 350\,ms to transfer the atoms into the 1560\,nm intra-cavity optical lattice with a predefined depth. The number of atoms transferred to the intracavity optical lattice is determined non-destructively by measuring the dispersive shift, $\omega_\mathrm{d}$, of the cavity by sweeping the frequency of a weak probe beam over the cavity resonance in 3 ms and recording the cavity transmission. 

A fixed atom number is maintained run-to-run using an FPGA which triggers the experiment once a set value is reached.  Explicitly, the cavity probe beam is set to a fixed frequency slightly less than the maximum dispersive shift. We then monitor the cavity transmission using the SPCM and record the count rate with the FPGA.  As the atoms are lost due to background collisions, the cavity is moved into resonance with the probe beam, increasing the output photon count rate. When the count rate reaches a preset threshold, the FPGA is triggered and clocks out the rest of the programmed experiment time sequence.  Strictly speaking, it is the dispersive shift that is fixed, and the accuracy of the atom number is limited by spatial averaging over the thermal distribution of atoms.  This procedure provides a high degree of repeatability in the experiment with evaporation ensuring a well defined temperature relative to the depth of the intracavity $1560\,\mathrm{nm}$ optical lattice, and triggering off a set dispersive shift then fixes a well defined atom number.  Remaining variation in the dispersive shift can be further reduced by post-selection, as it is also measured \emph{in situ} both immediately after the FPGA has been triggered and after the experiment is completed.

In the experiment, all lasers are referenced to a high finesse transfer cavity with a linewidth of $\sim 50\,\mathrm{kHz}$ at both $780\,\mathrm{nm}$ and $1560\,\mathrm{nm}$.  In addition, the experiment cavity is locked to the $1560\,\mathrm{nm}$ laser.  This allows all laser detunings to be accurately set relative to the empty experiment cavity resonance at $780\,\mathrm{nm}$.  The two laser fields in \fref{fig:transitiondiag} are obtained from a single laser using sidebands generated from a wide-band electro-optic modulator (EOM). Hence $(\omega_\mathrm{s}-\omega_\mathrm{r})/2$ is set by the driving frequency of the EOM and $\omega_c-(\omega_\mathrm{r}+\omega_\mathrm{s})/2$ is the detuning of the carrier with respect to the empty cavity. Complete specification of the model parameters then requires a measurement of $\omega_1$ and a characterization of $\lambda_\mathrm{r,s}$ in addition to the \emph{in situ} measurement of the dispersive shift.  

To characterize the coupling strengths $\lambda_\mathrm{r,s}$ as a function of power, we note that, when $\lambda_\mathrm{s} = 0$, the Hamiltonian reduces to a Tavis-Cummings interaction.  Weak probing of the cavity then provides an avoided crossing with a splitting that is determined by $\lambda_\mathrm{r}$.  Fitting the cavity transmission as a function of both the probe detuning with respect to the cavity and the cavity detuning with respect to the Raman resonance then allows us to extract both the coupling strength $\lambda_\mathrm{r}$ and the splitting $\omega_1$.  The splitting is shifted by the differential AC stark shift from the coupling beam, but this can be inferred from the measured dispersive shift, $\omega_\mathrm{d}$, the measured value of $\lambda_\mathrm{r}$, and accounting for thermal averaging.

\section{Dicke model threshold}
With the model parameters fully characterized, we can explore the expected threshold behavior.  As $\lambda$ is increased, a phase transition occurs at a threshold given by \cite{dimer2007proposed},
\begin{equation}
\lambda_c=\frac{1}{2}\sqrt{\frac{\omega_0}{\omega}\left(\omega^2+\kappa^2\right)}.
\label{Eq:thresholdDimer}
\end{equation}
This equation does not include the effect of $\delta$ or decoherence.  A small, non-zero $\delta$ simply displaces the value of $\omega$ and decoherence can be  accounted for by considering a decay rate $\gamma$ of the collective spin \cite{torre2016dicke,gelhausen2016many, bohnet2012relaxation}.  The result is the more general expression
\begin{equation}
\lambda_c=\frac{1}{2}\sqrt{\frac{\omega_0^2+\gamma^2}{-2\langle \sigma_\mathrm{z}\rangle\omega_0}\frac{(\omega-\delta/2)^2+\kappa^2}{\omega-\delta/2}},
\label{Eq:threshold}
\end{equation}
where $\langle \sigma_\mathrm{z}\rangle$ is the expectation value of the initial spin normalized by the number of atoms and accounts for imperfect initial state preparation.

As discussed in Appendix A, inhomogeneous broadening arising from the velocity distribution of the atoms can also modify the threshold in a manner qualitatively similar to \eref{Eq:threshold}.  Consideration of inhomogeneous broadening results in the threshold
\begin{equation}
\lambda_c=\sqrt{\frac{\sqrt{2}\gamma_\mathrm{d}\left[\left(\omega-\delta/2\right)^2+\kappa^2\right]}{8\left(\omega-\delta/2\right) F\left(\frac{\omega_0}{\sqrt{2} \gamma_\mathrm{d}}\right)}},
\label{Eq:threshold_inhomogeneous}
\end{equation}
where $F(x)$ is the Dawson function and $\gamma_\mathrm{d}$ is the rms Doppler shift.  This equation reduces to \eref{Eq:threshold} in the limit $\left(\gamma_\mathrm{d},\gamma,\langle \sigma_\mathrm{z}\rangle\right)\rightarrow \left(0,0,-0.5\right)$.

To determine the threshold experimentally, a fixed number of atoms is prepared in the $F=1$ level as described in the previous section.  For each run, after the FPGA trigger and \emph{in situ} dispersive shift measurement,  the laser fields are switched on at low power and then ramped over 3 ms to a final maximum power while monitoring the output from the cavity using an SPCM. The threshold is inferred from the point of the ramp at which the photon count reaches a preset value (10 counts/5$\mu \mathrm{s}$, to be well above the background).

\fref{threshold1}  show the measured threshold as a function of $\omega_0$ for fixed values of $\omega$.  For each graph we give a curve (black) given by \eref{Eq:thresholdDimer}, a fit (gray) to \eref{Eq:threshold} using $\gamma$ as the fitting parameter, and a theoretical plot (red/blue/green colour plot) accounting for the effects of inhomogeneous broadening as given by \eref{Eq:threshold_inhomogeneous}. In the experiment, two obvious mechanisms for decoherence are spontaneous emission and collisions. Spontaneous emission rates are proportional to $\lambda^2$ \cite{Zhiqiang2016} and we estimate values of a few hundred per second over the range of experimental values explored. These are too small to account for the fitted values of $\gamma$ as is the collision rate which is estimated to be $\sim 1000/\mathrm{s}$ for \fref{threshold1}. 
Alternatively, Doppler shifts appearing in \eref{eq:Doppler} can be expected to play a role, particularly for smaller values of $\omega_0$.  For the data in \fref{threshold1} we estimate rms Doppler shifts of $\gamma_\mathrm{d}=2\pi\times 59\,\mathrm{kHz}$.  The model used to derive \eref{Eq:threshold_inhomogeneous} explicitly includes Doppler shifts arising from the motion of the atoms and captures the qualitative behaviour of the observed thresholds .

\begin{figure}
 \includegraphics[width=0.5\textwidth]{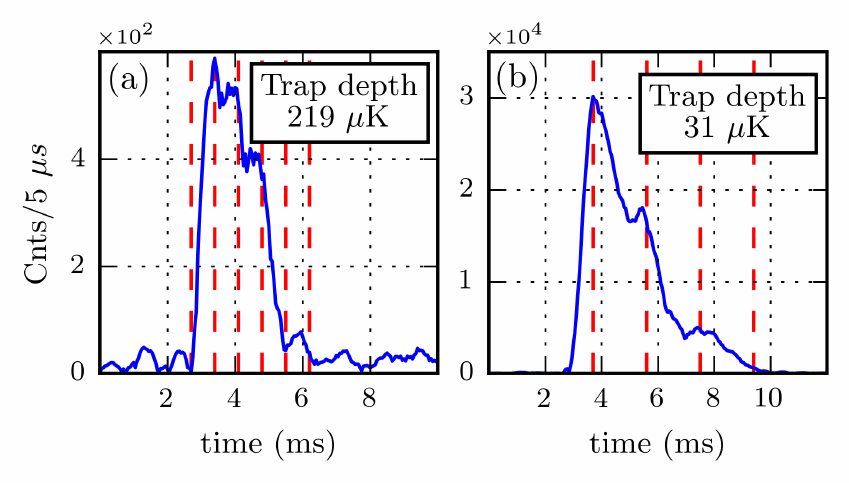}
 \caption{Typical cavity outputs observed using heterodyne detection. The parameters $\omega=2 \pi \times 100 \kHz$ and  $\omega_0=2 \pi \times 215 \kHz$ for both (a) and (b).  The trap depths are fixed at 219 $\mu$K  (a) and 31 $\mu$K (b).  Red dash lines are spaced at T/2 where T is the period of harmonic oscillation in the direction of the probe laser for the respective trap depths. }
\label{Output1}
\end{figure}

Trap depth also has a pronounced effect on output intensity from the cavity during the superradient phase. In \fref{Output1} (a) and (b), typical output signals are presented for the trap depths of $219\,\mathrm{\mu K}$ and $31\,\mathrm{\mu K}$ respectively.  These signals are derived from a heterodyne detection setup in order to avoid saturation of the SPCM.  Aside from the overall scale of the output signal, it is worth noting the output duration.  When including a decay mechanism for the collective spin, the resulting semiclassical equations predict a finite pulse duration with a timescale given by $1/\gamma$.  For the outputs shown in \fref{Output1}, this timescale is more consistent with those associated with the spontaneous emission and collisions, than with the values of $\gamma$, inferred by fitting the threshold data to \eref{Eq:threshold}, or $\gamma_\mathrm{d}$.  This suggests that the Doppler broadening is an important factor in reaching the transition but less so in the subsequent dynamics.  However apparent in the outputs are marked plateaus.  The vertical lines given in the plots are separated by half the period of oscillation associated with motion in the direction of the probe.  This suggests that periodic motion of the atoms may play a role in the dynamics beyond the phase transition.
\begin{figure}
\includegraphics[width=0.5\textwidth]{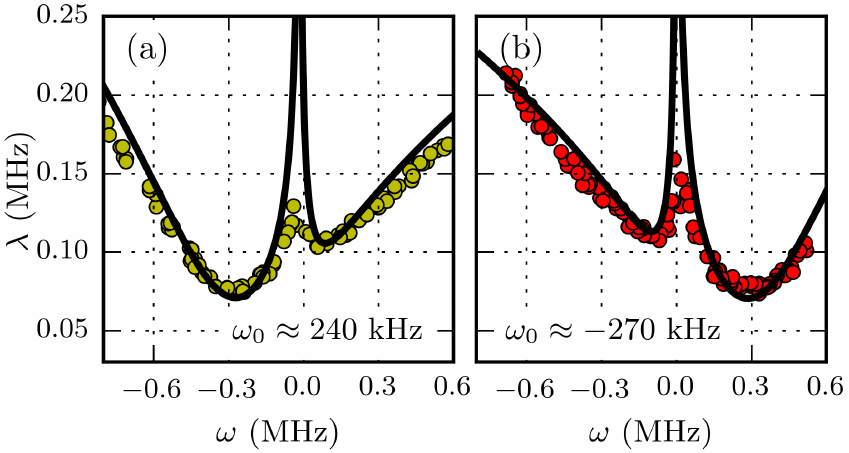}
\caption{Measured critical coupling as function of $\omega$ for (a) $\omega_0>0$  and (b) $\omega_0<0$ . Solid black lines show the critical coupling strength calculated  \emph{ab initio} from the theory given in Appendix A.}
\label{threshold3}
\end{figure}
\begin{figure}
 \includegraphics[width=0.5\textwidth]{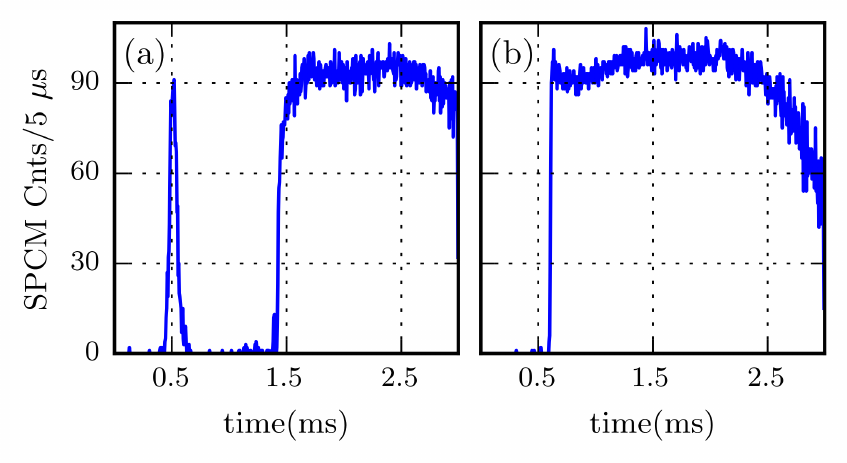}
 \caption{Typical cavity output pulses in the parameter regimes, (a) $\omega \omega_0<0$ and (b) $\omega \omega_0>0$.}
\label{pulses}
\end{figure}
\begin{figure}
 \includegraphics{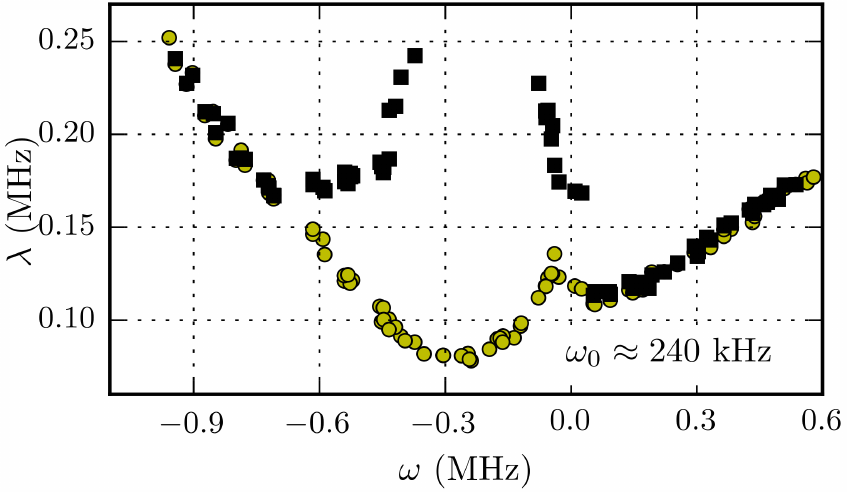}
 \caption{The yellow dots are the same as shown in \fref{threshold3}(a) where the threshold is determined from the first onset of light output from the cavity. The black squares are the thresholds extracted from the same data but neglecting the initial transient pulse, such as shown in \fref{pulses}(a), if one occurs. }
\label{Double}
\end{figure}

The experimental implementation allows both positive and negative values of $\omega$ and $\omega_0$.  \fref{threshold3} shows the obtained threshold with $\omega$ varying over both positive and negative values for a fixed positive value of  $\omega_0>0$  \footnote{Strictly speaking $\omega_0$ is not fixed as it depends on the differential dispersive shift.  However this only changes the value of $\omega_0$ by $\sim 10\,\mathrm{kHz}$ for the range of parameters explored.}. Changing the sign of both $\omega$ and $\omega_0$ amounts to changing the sign of both laser detunings from their respective cavity-assisted Raman resonances and is of no physical consequence.  The relative sign of $\omega$ and $\omega_0$, however, determines the interpretation of the initial state with the atoms pumped to $F=1$.  When $\omega$ and $\omega_0$ have the same sign, the initial state is given by $\ket{N/2,-N/2}$ which is a stable ground state of the Dicke-Hamiltonian below threshold.  When $\omega$ and $\omega_0$ have opposite signs, the initial state corresponds to $\ket{N/2, N/2}$.  This can be understood by considering a unitary transformation $U=\exp(-i \pi J_y)$ corresponding to a rotation about the $y$ axis by $\pi$.  The transformation effectively flips the sign of $\omega_0$ in Hamiltonian and the eigenvalue of $J_\mathrm{z}$ for the initial state.

\begin{figure*}
\includegraphics{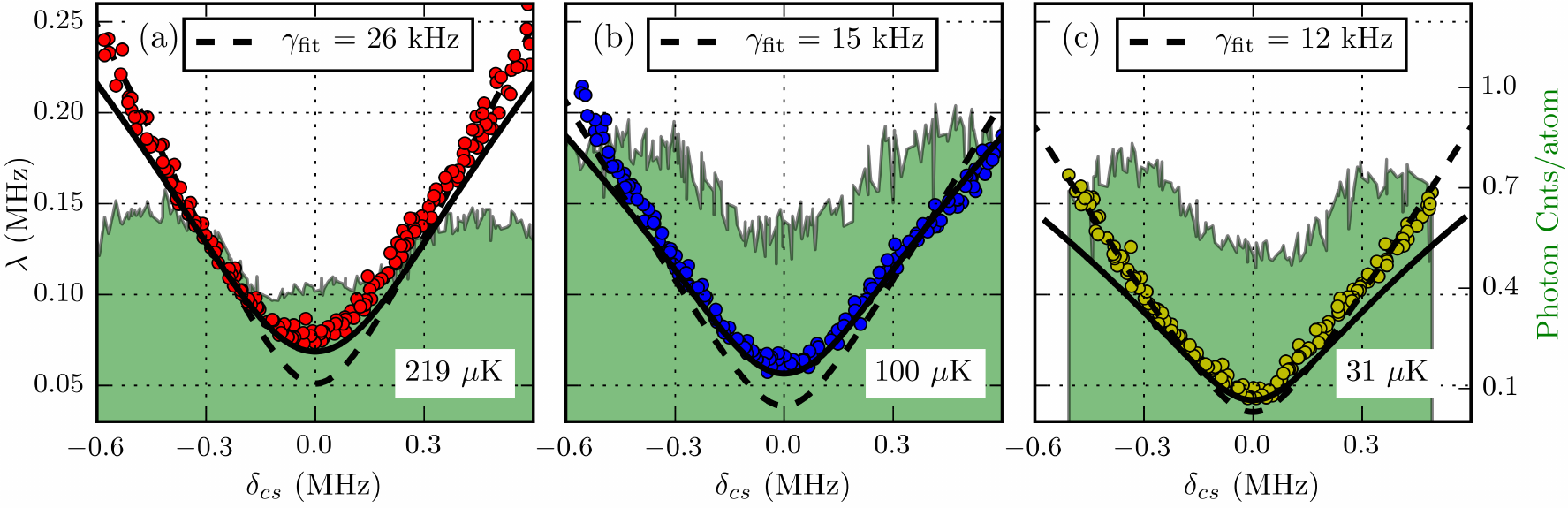}
\caption{Threshold for cavity output pulse for a single beam as function of detuning from the cavity-assisted Raman resonance.  Data in (a-c) is shown for three different traps depths: 219 $\mu$K, 100 $\mu$K, and 31 $\mu$K respectively. Black dash curves are fits to the experiment data points but only for $|\delta_\mathrm{cs}| > 250\kHz$ using \eref{eq:single_theory} with $\gamma$ as a free parameter. The rms Doppler broadening due to thermal motion for the respective plots are estimated to be $2 \pi \times [59, 40, 22] \kHz$. Black curves are obtained considering only the corresponding inhomogeneous broadenings as discussed in Appendix A using the experimentally determined Doppler broadening.}
\label{SingleThreshold}
\end{figure*} 

The relative sign of $\omega$ and $\omega_0$, or equivalently the initial state, has a clear signature in the output of the cavity as shown in \fref{pulses}, with (a) and (b) representing typical cavity outputs corresponding to $\omega>0$ and $\omega<0$, respectively, for the threshold measurements shown in \fref{threshold3}.  When starting in the proper stable state of the Dicke-model, that is $\omega>0$, the cavity output is characterized by a single pulse as expected. When starting in the unstable state, the expected cavity output is preceded by a smaller, much shorter pulse as in \fref{pulses}(a).   We interpret this short pulse as a single beam effect in which the atoms are transferred from the unstable state to the other.  In this case the threshold given in \fref{threshold3} is based on the appearance of the first pulse.  However we can define two such thresholds: one for the first pulse and one for the second.  In \fref{Double} we give an analysis showing the two thresholds with the data points in black obviously showing the threshold for the second pulse.  Ideally the first pulse would perfectly transfer the atoms from one state to the other which would provide a more symmetric plot as expected from theory.  The lack of symmetry indicates this is not the case.  Additionally, as $|\omega|$ increases, the two pulses appear closer together and eventually coalesce. This is because the Dicke model threshold eventually becomes smaller than the single beam threshold, as shown in \fref{Both}.  Note that the theoretical model given in Appendix A correctly captures the threshold behaviour across the entire parameter regime. 
 
\section{Single beam threshold}
Given the interpretation of the short duration pulse before superradiance, it is of interest to explore the single beam case.  In Fig.~\ref{SingleThreshold} we plot the threshold for this transient pulse with a single beam as a function of $\delta_\mathrm{cs}=-(\omega_0+\omega)$. For this case, a non-zero threshold occurs in the presence of decoherence \cite{bohnet2012relaxation}, in which case there must be sufficient driving to overcome decay of the collective spin.  The semiclassical equations can be used to determine a threshold and we obtain the expression:
\begin{equation}
\lambda_\mathrm{single} \ge \sqrt{\frac{\gamma\kappa}{-2 \langle \sigma_\mathrm{z} \rangle} \left( 1+\left(\frac{\delta_\mathrm{cs}}{\gamma+\kappa}\right)^2\right)},
\label{eq:single_theory}
\end{equation}
where $\langle \sigma_\mathrm{z} \rangle$ is the initial value which accounts for imperfect state preparation.

When fitting the above equation to the data, it is found that fits typically underestimate the threshold for larger values of $\delta_\mathrm{cs}$ and overestimate otherwise. If Doppler broadening were important, it would be reasonable to conjecture that it would be less so for larger values of $\delta_\mathrm{cs}$.  Hence, in Fig.~\ref{SingleThreshold} fits are given which exclude the points for which  $|\delta_\mathrm{cs}| \leq 250\kHz$.  Constrained in this way, the fits are notably better for lower temperatures and hence lower Doppler broadening.  Again, as in the Dicke-model case, the fitted decay rate is well above that expected from collisions and spontaneous emission but notably diminishes with temperature.  As shown in Appendix A, inhomogeneous Doppler broadening can also give rise to a non-zero threshold in the single beam case. Thresholds derived from that theory using the experimentally determined Doppler broadening are also included in Fig.~\ref{SingleThreshold} and are seen to better explain the data at least for small values of $|\delta_\mathrm{cs}|$.

\begin{figure}
 \includegraphics{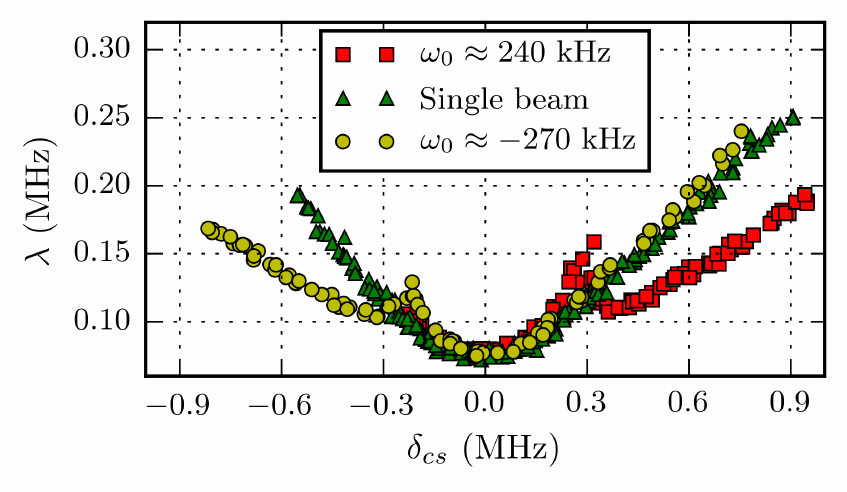}
 \caption{Dicke model threshold data from \fref{threshold3} (yellow dots and red squares) recast as a function of $\delta_\mathrm{cs}$ for direct comparison to the single beam threshold data (green triangles).}
\label{Both}
\end{figure}

It is also of interest to directly compare the single beam case to the results of the previous section.  To this end we determine the Dicke-model threshold as a function of $\delta_\mathrm{cs}$ for the two cases in which $\omega_0$ is fixed at around $\pm 2\pi\times 250\,\mathrm{kHz}$.  In Fig.~\ref{Both}, we plot the thresholds for these two cases, along with the single beam case.  The overlap in the region between around $\pm  2\pi\times 250\,\mathrm{kHz}$ is consistent with the previous interpretation that the first of the two pulses seen in this region is a single beam effect.
\section{Cavity transmission spectrum}
\begin{figure*}
\includegraphics[width = \columnwidth, trim = {0cm 2cm 0cm 4cm}, clip]{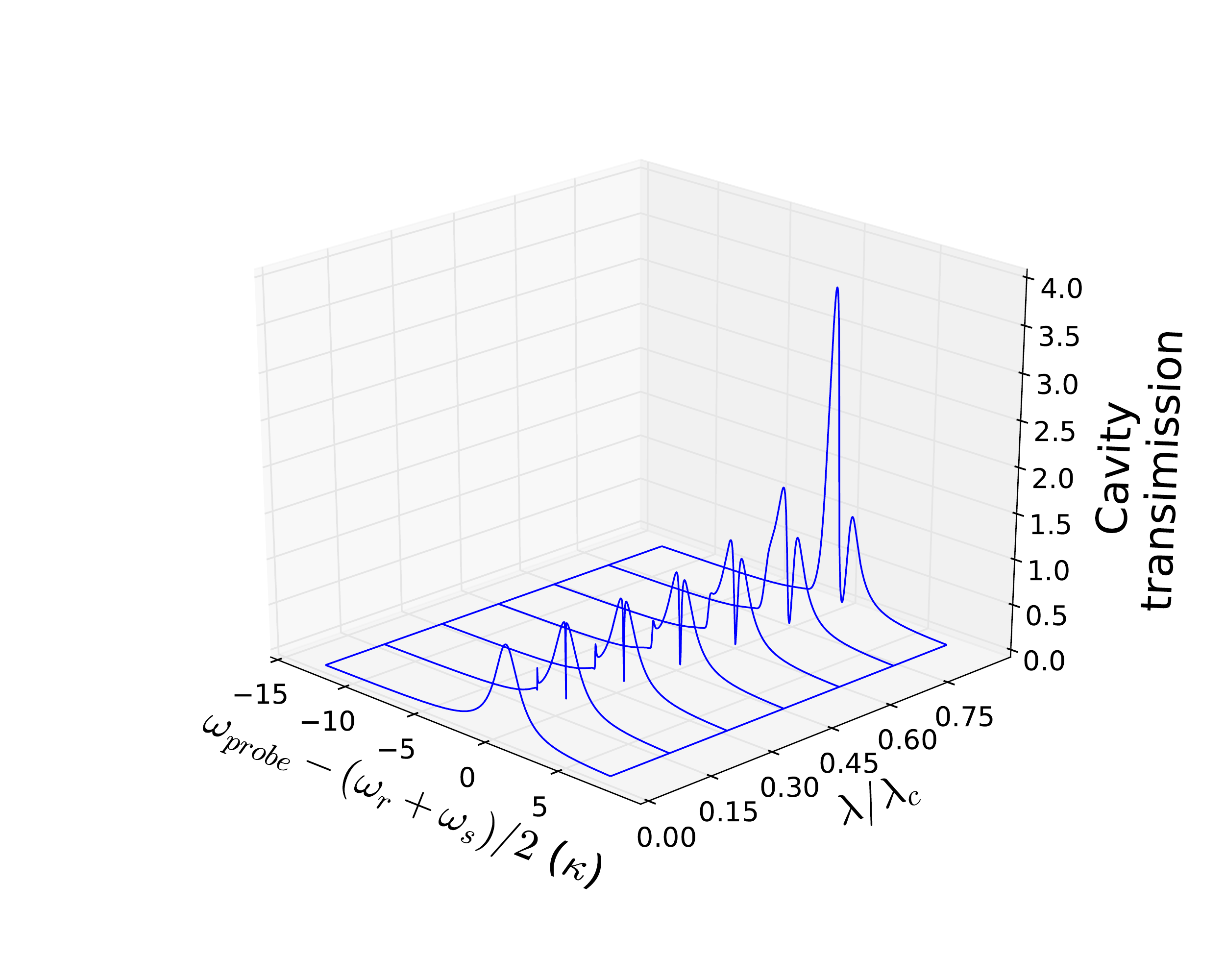}
\includegraphics[width = \columnwidth, trim = {0cm 1.7cm 0cm 4cm}, clip]{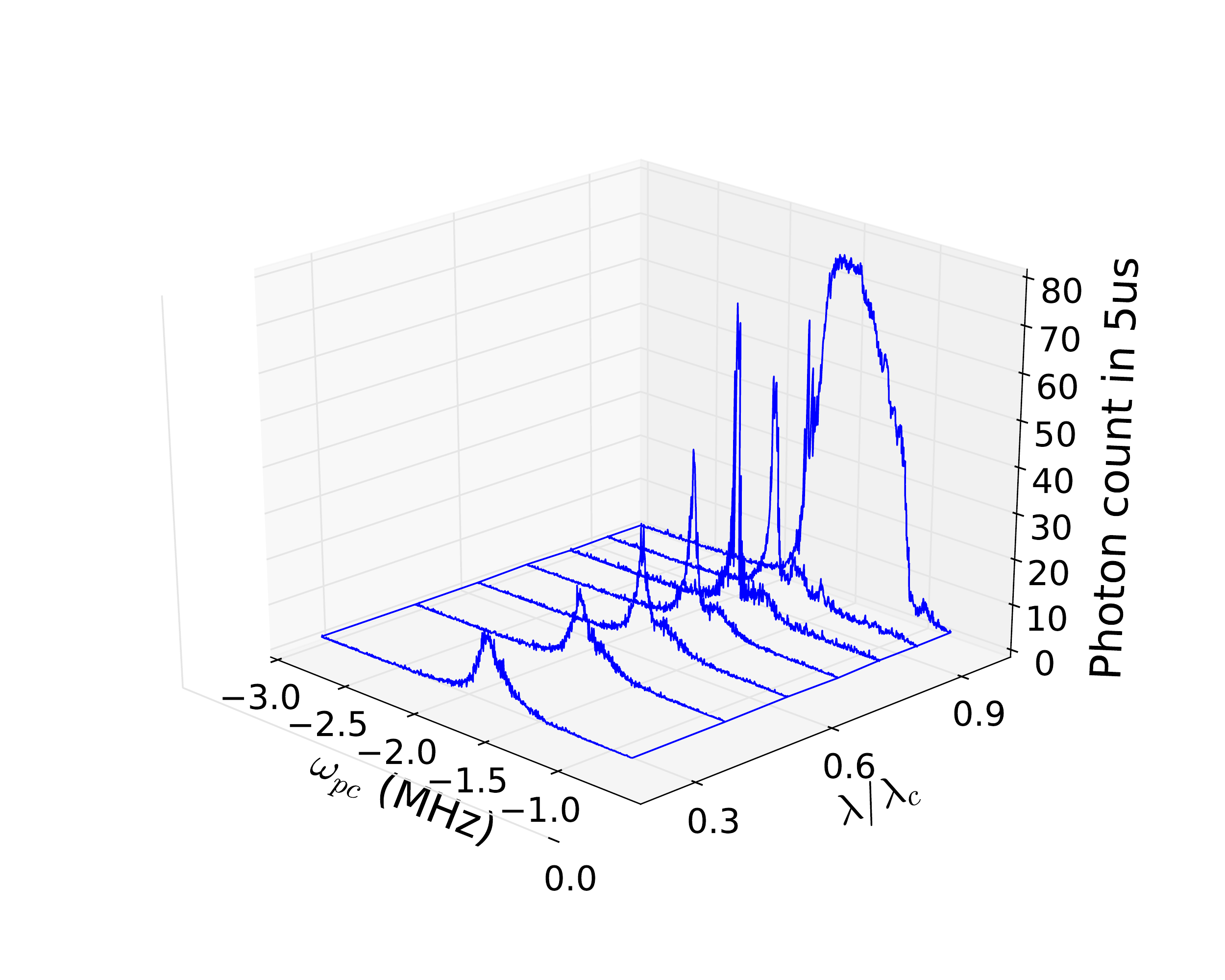}
\caption{(a) Theoretical transmission spectrum, calculated using the result of ref. \cite{dimer2007proposed}. (b) Measured transmission spectrum for $\omega = 100\,\text{kHz} = \omega_0$, and the dispersive shift $\omega_\mathrm{d} = -1.7\,$MHz. The threshold power is about 1.3\,mW. Above the critical coupling, the detector is saturated by the superradiant pulse, as can be seen in the final trace, and the transmission spectrum cannot be measured using the SPCM.}
  \label{fig:transmissionspec}
\end{figure*}

The improvement in stability and repeatability of the experiment has allowed us to verify the cavity transmission spectrum near the critical coupling as predicted in \cite{dimer2007proposed}. According to ref. \cite{dimer2007proposed}, as the coupling strength is increased close to the critical coupling, and the cavity is simultaneously probed with a weak classical beam, a peak is expected in the transmission at the average frequency of the two coupling lasers, $(\omega_\mathrm{s} + \omega_\mathrm{r}) / 2$. To observe this, the power of the two coupling beams is first raised to a value close to the measured threshold. The transmission spectrum is then measured by sweeping the frequency of the cavity probe beam and measuring the output photons. The results are plotted in Fig. \ref{fig:transmissionspec} along with the theoretical prediction. The larger peak in the measured transmission spectrum appears at smaller $\lambda / \lambda_c$ compared to the theoretical prediction, this is possibly due to the fact that the cavity probe beam is not sufficiently weak as is assumed in the theoretical analysis. The intracavity field therefore brings $\lambda$ above the critical coupling at a lower Raman beam power. This power of the cavity probe beam is, however, necessary to ensure a good signal to noise ratio in the transmission spectrum measurement. 

\section{Co-propagating vs counter-propagating}
\begin{figure}
 \includegraphics[width = \columnwidth]{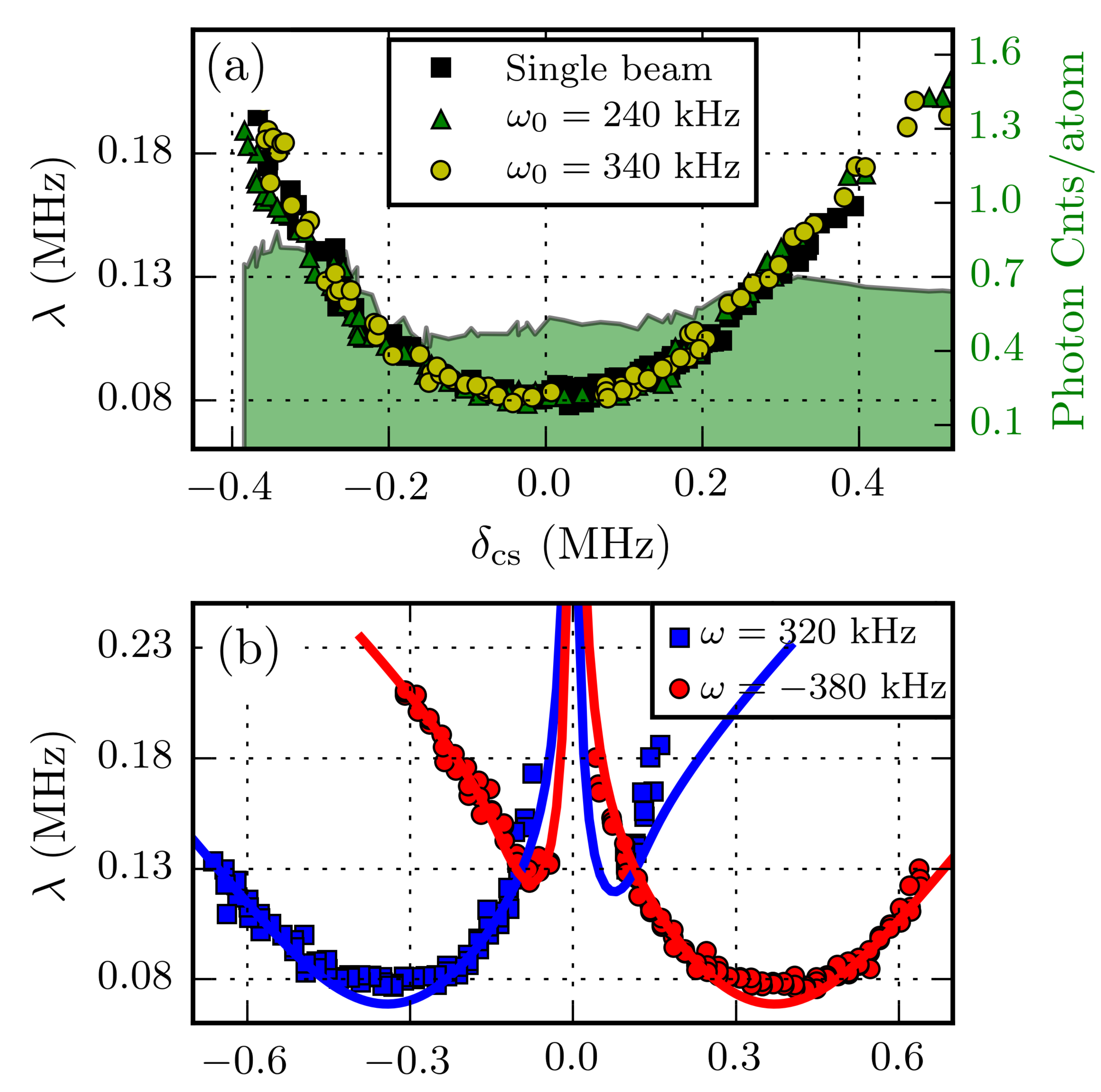}
 \caption{(a) The experiment in the co-propagating configuration with (green triangles) $\omega_0$ = 2$\pi \times $ 240 kHz and (yellow dots) $\omega_0$ = 2$\pi \times$ × 340 kHz. Single beam data also shown for comparison. The solid green background is the average number of photons lost from the cavity per atom for the co-propagating data run (green triangles). (b) Threshold in co-propagating configuration as function of $\omega_0$ for two values of $\omega$ illustrating the narrow gap that occurs when $\omega_0 \approx 0$. Solid lines show the critical coupling strength calculated  \emph{ab initio} from the theory given in Appendix A.}
\label{Comparison1}
\end{figure}

From Sect.~\ref{Sect:theory}, it is clear that the co-propagating case, as used in \cite{baden2014realization}, does not support a simple Dicke-model interpretation.  We have extensively reinvestigated this geometry over a wide range of parameters and, in most cases, the second beam plays no significant role.  We illustrate this in Fig.~\ref{Comparison1}(a) in which threshold data, as a function of the single beam detuning $\delta_\mathrm{cs}$, is shown for both a single beam and co-propagating beams with two different values of $\omega_0$.  This highlights the independence of $\omega_0$ and hence the presence of the other beam.

The size of the observed cavity output pulse in the co-propagating versus counter-propagating configuration further supports this conclusion. In the single beam case, the complete population transfer of $N$ atoms can generate at most $N$ photons in the cavity.  By integrating the total number of counts in the output pulse, and accounting for the quantum efficiency of detecting an excitation in the cavity, we can estimate the average number of photons scattered per atom in the duration of the pulse. In all the data for the single beam data and co-propagating configuration (both from \cite{baden2014realization}  and repeated experiments) the output pulse is typically short in duration ($\sim$ 100 $\mu$s) and the average number of photons scattered per atom is strictly $< 1$, consistent with a (partial) population transfer in both cases. 

As illustrated in \fref{Output1}(b), outputs in the counter-propagating configuration are sustained for several milliseconds with more than a hundred photons per atom scattered into the cavity. This strongly supports the conclusion that a quasi-stationary superradient state consistent with the Dicke-model phase transition is observed in the counter-propagation configuration. The output pulses in the co-propagating configuration as observed in \cite{baden2014realization} on the other hand, are identical to those observed with only a single beam and thus do not support the conclusion of \cite{baden2014realization} that a Dicke-model simulation had been realized.

There is only one small parameter regime in which the second beam plays a significant role.  This occurs when $\omega_0\approx 0$ or equivalently when the laser detunings from their respective cavity-assisted Raman resonances are equal. The observed threshold as a function of $\omega_0$ for a fixed value of $\omega=2\pi\times 320 \kHz$ and $\omega=2\pi\times -380 \kHz$  are given in Fig.~\ref{Comparison1}(b) which clearly shows that the second beam effectively blocks the cavity-assisted Raman transfer when $\omega_0\approx 0$.  This phenomenon is adequately captured by the threshold model given in Appendix A, as illustrated by the solid curves given in \fref{Comparison1}(b).  

\section{Conclusion}
In this paper we have explored the use of cavity assisted Raman transitions for the simulation of a Dicke-model Hamiltonian.  Thresholds for the single beam, co-propagating and counter-propagating configurations have been explored and we have demonstrated that phase-matching in the counter-propagating configuration is critical for a Dicke-model interpretation. In the counter-propagating case, the prediction of the cavity transmission spectrum close to the phase transition was also verified.  A theoretical description that accounts for the beam geometry used in the experiment indicates the potential role of motional effects and evidence of these effects are seen in the experimental results. It would be of interest to develop the theory further: to fully explore the consequences of motional coupling, how it effects the observed cavity outputs, and how the atomic motion is influenced by the onset of super-radiance.

\onecolumngrid
\section*{Acknowledgements}
This research is supported by the National Research Foundation, Prime Ministers Office, Singapore and the Ministry of Education, Singapore under the Research Centres of Excellence programme (partly under grant No. NRF-CRP12-2013-03). S. J. Masson and A. S. Parkins acknowledge support from the Marsden Fund of the Royal Society of New Zealand (Contract No. UOA1328). 

\section*{APPENDIX A: Inhomogeneous broadening effects}
Here we consider the effects of inhomogeneous broadening on the observed thresholds in the single beam and two beam configurations.  We assume the internal atomic and cavity dynamics are much faster than timescales governing external motion, and treat the position as a classical parameter.  The Hamiltonian is then
\begin{equation}
H = \omega a^\dagger a + \omega_0 \sum_{j=1}^N \sigma_{z,j} + \frac{\lambda_r}{\sqrt{N}}\sum_{j=1}^N \left(e^{-i\phi_j}a\sigma_j^\dagger+e^{i\phi_j}a^\dagger\sigma_j\right)+ \frac{\lambda_s}{\sqrt{N}}\sum_{j=1}^N \left(e^{\pm i\phi_j}a^\dagger\sigma_j^\dagger+e^{\mp i\phi_j}a\sigma_j\right),
\end{equation}
where the phases for each atom are given by $\phi_j=\mathbf{k}\cdot\mathbf{r}_j(t)=k r_j(t)$.  The upper and lower signs in the last summation correspond to the co- and counter-propagating beam configurations respectively.  Since we are considering threshold, we use the Holstein-Primakoff approximation $\sigma_j\approx b_j$, which is valid provided $b_j^\dagger b_j\ll1$.  The corresponding quantum Langevin equations are then 
\begin{equation}
\dot{a}=-(\kappa+i\omega) a-\frac{i \lambda_r}{\sqrt{N}}\sum_{v} e^{ik r_j(t)} b_j-\frac{i\lambda_s}{\sqrt{N}} \sum_{v} e^{\pm i k r_j(t)} b_j^\dagger+\sqrt{2\kappa} a_\mathrm{in}(t)
\end{equation}
and
\begin{equation}
\dot{b}_j=-i\omega_0 b_j-\frac{i \lambda_r}{\sqrt{N}} e^{-ik r_j(t)} a-\frac{i\lambda_s}{\sqrt{N}} \sum_{v} e^{\pm i k r_j(t)} a^\dagger.
\end{equation}
Substituting the formal solution 
\begin{equation}
b_j(t)=e^{-i\omega_0 t} b_j(0)-\frac{i \lambda_r}{\sqrt{N}} \int_0^t e^{-ik r_j(t')} e^{-i \omega_0 (t-t')}a(t') dt' -\frac{i\lambda_s}{\sqrt{N}}  \int_0^t e^{\pm i k r_j(t')} e^{-i\omega_0 (t-t')}a^\dagger(t')dt',
\end{equation}
into the equation for $\dot{a}$ gives
\begin{multline}
\dot{a}=-(\kappa+i\omega) a-\frac{i \lambda_r}{\sqrt{N}}\sum_{j} e^{ik r_j(t)} e^{-i\omega_0 t} b_j(0)-\frac{i \lambda_s}{\sqrt{N}}\sum_{j} e^{\pm ik r_j(t)} e^{i\omega_0 t} b_j^\dagger(0)+\sqrt{2\kappa} a_\mathrm{in}(t)\\
-\frac{\lambda_r^2}{N}\left(\sum_{j}\int_0^t dt' e^{i k(r_j(t)-r_j(t'))}e^{-i\omega_0(t-t')}a(t')\right)+\frac{\lambda_s^2}{N}\left(\sum_{j}\int_0^t dt' e^{\pm i k(r_j(t)-r_j(t'))}e^{i\omega_0(t-t')}a(t')\right)\\
-\frac{\lambda_r\lambda_s}{N}\sum_{j} \left(\int_0^t dt' e^{i k(r_j(t) \pm r_j(t'))} e^{-i\omega_0(t-t')}a^\dagger(t')-\int_0^t dt' e^{\pm i k(r_j(t) \pm r_j(t'))} e^{i\omega_0(t-t')}a^\dagger(t')\right).
\end{multline}
Considering expectation values with $\langle b_v(0)\rangle=\langle b_v^\dagger(0)\rangle=\langle a_\mathrm{in}(t)\rangle=0$ gives
\begin{multline}
\label{aeq}
\langle \dot{a}\rangle =-(\kappa+i\omega) a\\
-\frac{\lambda_r^2}{N}\left(\sum_{j}\int_0^t dt' e^{i k(r_j(t)-r_j(t'))}e^{-i\omega_0(t-t')}\langle a(t')\rangle \right)+\frac{\lambda_s^2}{N}\left(\sum_{j}\int_0^t dt' e^{\pm i k(r_j(t)-r_j(t'))}e^{i\omega_0(t-t')}\langle a(t')\rangle \right)\\
-\frac{\lambda_r\lambda_s}{N}\sum_{j} \left(\int_0^t dt' e^{i k(r_j(t) \pm r_j(t'))} e^{-i\omega_0(t-t')}\langle a^\dagger(t')\rangle -\int_0^t dt' e^{\pm i k(r_j(t) \pm r_j(t'))} e^{i\omega_0(t-t')}\langle a^\dagger(t')\rangle\right).
\end{multline}
Note that only the cross terms proportional to $\lambda_r \lambda_s$ depend on the relative propagation direction of the beams.  Since the cavity and atomic evolution is much faster than the external motion, we take $r_j(t)\approx r_j+v_j t$ where $r_j$ and $v_j$ are now to be considered initial conditions.  In this case the position dependence drops out of the first two integrals and is only relevant for the other two in the co-propagating case.  The summations can be approximated by integrals of the form
\begin{equation}
\frac{1}{N}\sum_j f(r_j,v_j)=\frac{1}{2\pi \bar{v} \sigma}\iint dv dr \exp\left(-\frac{r^2}{2\sigma^2}\right)\exp\left(-\frac{v^2}{2\bar{v}^2}\right)f(r,v),
\end{equation}
where $\bar{v}$ is the rms velocity and $\sigma$ is the rms position.  The cross-terms do not contribution in the co-propagating case, since 
\begin{equation}
\frac{1}{\sqrt{2\pi\sigma^2}}\int_{-\infty}^{\infty} \exp\left(-\frac{r^2}{2\sigma^2}\right) e^{2ik r}dr=e^{-2(k\sigma)^2}\approx 0,
\end{equation}
for distributions relevant to the experiments.  Thus, in all cases we are left with terms of the form
\begin{equation}
\frac{1}{\sqrt{2\pi \bar{v}^2}}\int_{-\infty}^\infty \int_0^t \exp\left(-\frac{v^2}{2 \bar{v}^2}\right) e^{\pm i (\omega_0-kv)(t-t')}\langle \mathcal{O}(t')\rangle dt'dv=\int_0^t \exp\left(-\frac{\gamma_\mathrm{d}^2(t-t')^2}{2}\right) e^{\pm i\omega_0(t-t')}\langle \mathcal{O}(t')\rangle dt',
\end{equation}
where $\gamma_\mathrm{d}=k\bar{v}$.  The Laplace transform then gives
\begin{align}
\int_0^\infty e^{-s t}&\left(\int_0^t \exp\left(-\frac{\gamma_\mathrm{d}^2(t-t')^2}{2}\right) e^{\pm i\omega_0(t-t')}\langle \mathcal{O}(t')\rangle dt'\right)dt\\
&=\int_0^\infty \int_{t'}^\infty \exp\left(-\frac{\gamma_\mathrm{d}^2(t-t')^2}{2}\right) e^{-s t\pm i\omega_0(t-t')}\langle \mathcal{O}(t')\rangle dt dt'\\
&=\int_0^\infty \left(\int_{t'}^\infty \exp\left(-\frac{\gamma_\mathrm{d}^2(t-t')^2}{2}\right) e^{-(s\mp i\omega_0)(t-t')}dt \right)e^{-s t'}\langle \mathcal{O}(t')\rangle dt'\\
&=\int_0^\infty \left(\int_{0}^\infty \exp\left(-\frac{\gamma_\mathrm{d}^2\tau^2}{2}\right) e^{-(s\mp i\omega_0)\tau}d\tau \right)e^{-s t'}\langle \mathcal{O}(t')\rangle dt'\\
&=\frac{1}{\gamma_\mathrm{d}}\sqrt{\frac{\pi}{2}} \exp\left(z_\mp^2\right) \mathrm{erfc}(z_\mp) \langle \tilde{\mathcal{O}}(s)\rangle,
\end{align}
where $z_\pm=(s\pm i\omega_0)/(\gamma_\mathrm{d}\sqrt{2})$, $\langle \tilde{\mathcal{O}}(s)\rangle$ is the Laplace transform of $\langle {\mathcal{O}}(t)\rangle$, and $\mathrm{erfc}(z)$ is the complementary error function.  The Laplace transform of Eq.~\ref{aeq} then gives
\begin{multline}
(s+\kappa+i\omega)\langle\tilde{a}(s)\rangle-a(0)=-\frac{\lambda_r^2}{\gamma_\mathrm{d}}\sqrt{\frac{\pi}{2}} f(z_+) \langle \tilde{a}(s)\rangle+\frac{\lambda_s^2}{\gamma_\mathrm{d}}\sqrt{\frac{\pi}{2}} f(z_-) \langle \tilde{a}(s)\rangle\\
-\frac{\lambda_r\lambda_s}{\gamma_\mathrm{d}}\sqrt{\frac{\pi}{2}} f(z_+) \langle \tilde{a}^\dagger(s)\rangle+\frac{\lambda_r\lambda_s}{\gamma_\mathrm{d}}\sqrt{\frac{\pi}{2}} f(z_-) \langle \tilde{a}^\dagger(s)\rangle,
\end{multline}
where $f(z)=e^{z^2}\mathrm{erfc}(z)$ and $\lambda_r\lambda_s$ is set to zero in the co-propagating case.  It is convenient to scale $s,\kappa, \omega, \omega_0, \lambda_{r,s}$ and $a(0)$ by the factor $\gamma_\mathrm{d}\sqrt{2}$ to give the scaled form
\begin{equation}
(z+\bar{\kappa}+i\bar{\omega})\langle\tilde{a}(z)\rangle-\bar{a}(0)=-(\bar{\lambda}_r^2 f(z_+)-\bar{\lambda}_s^2 f(z_-))\sqrt{\pi} \langle \tilde{a}(z)\rangle\\
-\bar{\lambda}_r \bar{\lambda}_s\sqrt{\pi} (f(z_+)-f(z_-)) \langle \tilde{a}^\dagger(z)\rangle.
\end{equation}
If any of the poles of $\langle \tilde{a}(z) \rangle$ have a positive real part, the solution describes an amplification process, that is, a growth of the cavity field and transfer of atomic population.  The threshold is then determined by the smallest value of $\bar{\lambda}$ at which a pole crosses the imaginary axis.

For the single beam case ($\bar{\lambda}_r=0$), the Laplace transform solution is
\begin{equation}
\langle \tilde{a}(z) \rangle=\frac{\bar{a}(0)}{z_-+z_0-\bar{\lambda}^2\sqrt{\pi} f(z_-)}
\end{equation}
where $z_0=\bar{\kappa}+i(\bar{\omega}+\bar{\omega}_0)=\bar{\kappa}-i \bar{\delta}_{cs}$.  Since $z_-$ is simply a translation of $z$ along the imaginary axis, the threshold is also indicated by poles in the positive half-space of the $z_-$ plane.  The threshold can be found by the requirement that the denominator of $\langle \tilde{a}(z) \rangle$ is zero for $z_-=i y$.  This gives $\bar{\lambda}^2=\bar{\kappa} e^{y^2}/\sqrt{\pi}$ where $y$ satisfies
\begin{equation}
\bar{\delta}_{cs}=y+\frac{2 \bar{\kappa}}{\sqrt{\pi}}e^{y^2} F(y),
\end{equation}
in which $F(y)$ is the Dawson function defined by
\begin{equation}
F(y)=e^{-y^2}\int_0^y e^{x^2}dx.
\end{equation}
For $\bar{\delta}_{cs}=0$, $y=0$ giving the resonant threshold $\bar{\lambda}^2=\bar{\kappa}/\sqrt{\pi}$.  All other values can be found numerically.

In the co-propagating case $(\bar{\lambda}_r=\bar{\lambda}_s=\bar{\lambda}, \bar{\lambda}_r \bar{\lambda}_s\rightarrow0)$, the Laplace transform is
\begin{equation}
\langle \tilde{a}(z) \rangle=\frac{\bar{a}(0)}{z+\bar{\kappa}+i\bar{\omega}+\bar{\lambda}^2\sqrt{\pi} (f(z_+)-f(z_-))}.
\end{equation}
In the special case $\omega_0=0$, $z_+=z_-$ and there is no dependence on $\bar{\lambda}$.  Consequently no threshold exists.  For other values, the threshold can be found numerically.

In the counter-propogating case $\bar{\lambda}_s=\bar{\lambda}_r=\bar{\lambda}$ and the Laplace transform is
\begin{equation}
(z+\bar{\kappa}+i \bar{\omega})\langle \tilde{a}(z) \rangle-\bar{a}(0)=\bar{\lambda}^2\sqrt{\pi} (f(z_-)-f(z_+))(\langle \tilde{a}(z) \rangle+\langle \tilde{a}^\dagger(z) \rangle),
\end{equation}
with the corresponding equation for $\langle \tilde{a}^\dagger(z) \rangle$,
\begin{equation}
(z+\bar{\kappa}-i \bar{\omega})\langle \tilde{a}^\dagger(z) \rangle -\bar{a}^\dagger(0)=-\bar{\lambda}^2\sqrt{\pi} (f(z_-)-f(z_+))(\langle \tilde{a}(z) \rangle+\langle \tilde{a}^\dagger(z) \rangle).
\end{equation}
These two equations may be expressed in matrix form
\begin{equation}
\begin{pmatrix}
z+\bar{\kappa}+i \bar{\omega}-\bar{\lambda}^2\sqrt{\pi} (f(z_-)-f(z_+)) & -\bar{\lambda}^2\sqrt{\pi} (f(z_-)-f(z_+))\\
\bar{\lambda}^2\sqrt{\pi} (f(z_-)-f(z_+)) & z+\bar{\kappa}-i \bar{\omega}+\bar{\lambda}^2\sqrt{\pi} (f(z_-)-f(z_+))
\end{pmatrix}
\begin{pmatrix}
 \tilde{a}(z) \rangle\\ \tilde{a}^\dagger(z) \rangle
\end{pmatrix}=\begin{pmatrix}\bar{a}(0)\\\bar{a}^\dagger(0)\end{pmatrix}.
\end{equation}
Poles in $\langle \tilde{a}(z) \rangle$ are then roots of the determinant
\begin{equation}
(z+\bar{\kappa})^2+\bar{\omega}^2+2 i \bar{\omega} \bar{\lambda}^2\sqrt{\pi} (f(z_-)-f(z_+)).
\end{equation}
When $\omega \omega_0 < 0$, he threshold can be found numerically. When $\omega \omega_0 \ge 0$,roots of this expression first cross into the positive half-space of the $z$-plane along the real axis.  Hence we can set $z=0$ and solve for $\bar{\lambda}$.  This gives the expression
\begin{equation}
\bar{\lambda}=\sqrt{\frac{\bar{\omega}^2+\bar{\kappa}^2}{8\bar{\omega} F(\bar{\omega}_0)}}.
\end{equation}
Since we are considering behaviour below threshold, the non-linear term$\left(\delta a^\dagger a J_\mathrm{z}/N\right)$ in \eref{Eq:Hamiltonian} simply adds a small detuning $\left(-\delta/2\right)$ to $\omega$.

\twocolumngrid

\bibliographystyle{unsrt}

\bibliography{Dicke_SpinHalf}
\end{document}